\documentclass[%
 reprint,
 amsmath,amssymb,
 aps,
]{revtex4-2}
\usepackage{tipa}
\usepackage[utf8]{inputenc}
\usepackage{siunitx}
\usepackage{bm}
\usepackage{xcolor}
\usepackage{graphicx}
\usepackage{dcolumn}
\usepackage{bm}

\begin{document}

\preprint{APS/123-QED}

\title{Hybrid design of spectral splitters and concentrators of light for solar cells using iterative search and neural networks}

\author{Alim Yolalmaz$^{1, 2, 3,}$}
\email{alim.yolalmaz@metu.edu.tr}
 
\affiliation{
 $^{1}$Programmable Photonics Group, Department of Physics, Middle East Technical University, 06800 Ankara, Turkey\\
 $^{2}$Micro and Nanotechnology Program, Middle East Technical University, 06800 Ankara, Turkey\\
 $^{3}$The Center for Solar Energy Research and Applications (ODT\"{U}-G\"{U}NAM), Middle East Technical University, 06800 Ankara, Turkey
}%

\author{Emre Y\"{u}ce$^{1, 2, 3,}$}
 \email{eyuce@metu.edu.tr}
\affiliation{
  $^{1}$Programmable Photonics Group, Department of Physics, Middle East Technical University, 06800 Ankara, Turkey\\
 $^{2}$Micro and Nanotechnology Program, Middle East Technical University, 06800 Ankara, Turkey\\
 $^{3}$The Center for Solar Energy Research and Applications (ODT\"{U}-G\"{U}NAM), Middle East Technical University, 06800 Ankara, Turkey
}%

\date{\today}

\begin{abstract}
The need for optically multi-functional micro- and nano-structures is growing in various fields. Designing such structures is impeded by the lack of computationally low-cost algorithms. In this study, we present a hybrid design scheme, which relies on a deep learning model and the local search optimization algorithm, to optimize a diffractive optical element that performs spectral splitting and spatial concentration of broadband light for solar cells. Using generated data set during optimization of a diffractive optical element, which is a one-time effort, we design topography of diffractive optical elements by using a deep learning-based inverse design scheme. We show that further iterative optimization of the reconstructed diffractive optical elements increases amount of spatially concentrated and spectrally split light. Our joint design approach both speeds up optimization of diffractive optical elements as well as providing better performance at least 57\% excess light concentration with spectral splitting. The algorithm that we develop here will enable advanced and efficient design of multi-functional phase plates in various fields besides the application that we target in solar energy. The algorithm that we develop is openly available to contribute to other applications that rely on phase plates.

\end{abstract}

\maketitle


\section{\label{sec:level1}Introduction}

Enhanced control of propagation, diffraction, scattering, and interference of light using micro- and nano-structures plays an important role in tailoring light for a specific application\cite{Tittl2018, Zheng2015}. A phase plate can modulate direction, amplitude, phase, and polarization of the light or a combination of these properties which makes them superior to conventional optical structures\cite{Aieta2015, Wan2014, Aieta2012}. The multi-dimensional control provided by a phase plate yields to versatile performance increase in imaging\cite{Khorasaninejad2016}, holography\cite{ Mico2006a, Mico2006}, light focusing\cite{Yildirim2020}, spectral control\cite{Yolalmaz2020}, microscopy\cite{Bishara2010}, aberration correction\cite{Balli2020}, wavelength-multiplexing\cite{Ogura2001}, and solar energy\cite{Guen2021}. Although, the control on phase provides numerous advantages, the phase being sensitive at dimensions close to the wavelength of incident light brings extra challenges to overcome. The increased resolution requirement in designing a phase plate also increases the number of parameters to be optimized for an efficient control on light. A diffractive optical element (DOE) is a special type of phase plate that modulates light by using diffraction and interference with preferably minimal light scattering\cite{Stanley2016, Kim2013, Mohammad2014, Mohammad2016, Mojiri2013}.

Current research of solar cell technology focuses on improving performance of the solar cell materials. However, the theoretical conversion efficiency of the solar cells (30\%) for a single-junction is limited by spectrally bounded absorption of single-junction solar cells\cite{Shockley1961}. The small overlap between the emission spectrum of the sun and the absorption spectrum of a solar cell causes inefficient use of available energy. Solar energy can be more effectively converted to electricity using smart design schemes. Laterally arranged solar cells that are made from two different materials, produce increased energy output by enhancing spectral overlap between the absorption spectra of solar cells and the incident spectrum of the sunlight\cite{Stanley2016, Kim2013, Mohammad2014, Mohammad2016, Mojiri2013}. A DOE can spectrally disperse the sunlight and steer individual spectral band to the relevant target solar cell. We classify these multi-functional diffractive optical elements as spectral splitters and concentrators (SpliCons)\cite{Guen2021} to distinguish from DOEs that can only achieve spectral splitting. In this manuscript, we will use SpliCon for phase plates that achieve simultaneous spectral splitting and concentration and DOE for phase plates that can only achieve spectral splitting or concentration.

There are several design approaches to tune the structural parameters of the DOEs/SpliCons. These are direct binary search\cite{Seldowitz1987, Mohammad2018}, Gerchberg-Saxton\cite{Gerchberg1972, Wang2017, Vorndran2015}, and Yang-Gu algorithms\cite{Yang1994}, genetic optimization\cite{Johnson1993}, the local search optimization\cite{Yolalmaz2020a}, and gradient-based electromagnetic field optimization\cite{Xiao2016}. These design approaches for the DOEs/SpliCons have some bottlenecks which limit their wide-spread deployment. Moreover, the aforementioned probabilistic and gradient-based design approaches require immense computational resources. Compared to these computational tools, deep learning offers a fast inverse design path for the DOEs/SpliCons. Deep neural networks consist of multiple layers that connect input data to output data by revealing the correlation between the output and the input. Higher-level information within a data set, which is presented as weights of the layers, is captured, and thereby, complex network relations between the input data and the output data can be understood. Deep learning proves its adaptability and multi-operational behaviour in various tasks especially in photonics in designing nanophotonic devices\cite{Chen2020, Wiecha_2021}, anti-reflective surfaces\cite{Haghanifar2020}, and photonics systems\cite{Zhu2020, Jiang2019, Yolalmaz2021, Chugh2019}. Also, physics-informed neural networks can provide a better understanding of the underlying physics in various tasks\cite{Raissi2019}. Moreover, deep learning enables us to inverse-design a DOE/SpliCon within a short time with performance scores that are extremely close to the ground-truth values.

In this study, we present a hybrid design algorithm that combines deep learning and the local search optimization algorithm to speed up the optimization duration and enhance the DOE \& SpliCon topography. We created a training data set to develop a deep learning model through optimization of a DOE/SpliCon with the local search optimization algorithm, which is a one-time effort. Using the DOEs/SpliCons generated via the local search optimization algorithm we calculate spatial and spectral distribution of the intensity pattern via Fresnel-Kirchhoff diffraction integral at the target. Then we use deep learning-based optimization to reconstruct DOEs/SpliCons, which provides a much faster optimization. After that, we use hybrid optimization to further improve the performance of the DOEs/SpliCons for spectral splitting and spatial concentration of the broadband light. Our hybrid optimization procedure provides \textit{fast} and \textit{efficient} design of SpliCons for solar energy applications.

\begin{figure*}[!htb]
  \centering
  
   
   \includegraphics[width=180mm]{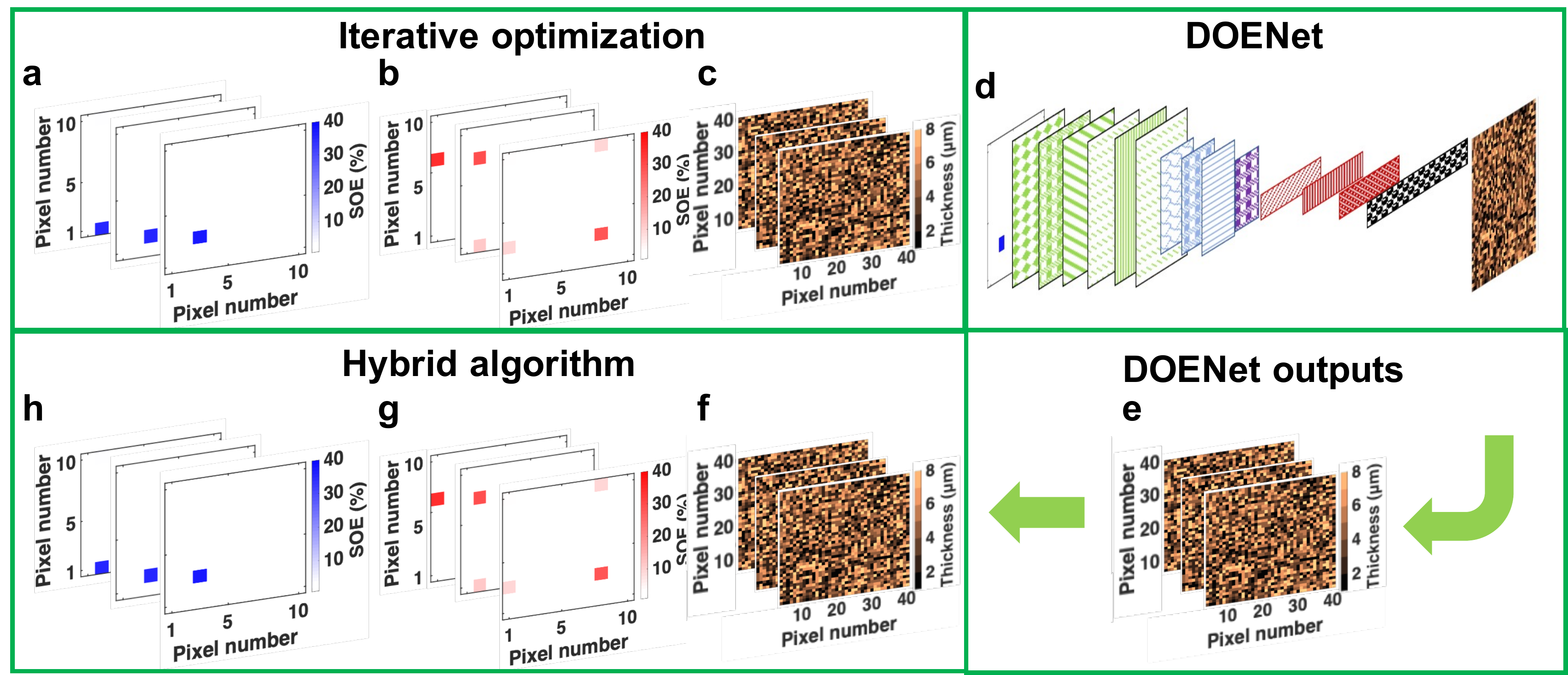}

  \caption{The flowchart of our hybrid algorithm. Input data sets which are SOE distributions at (a) 700 nm and (b) 1100 nm at the output plane for alternating target positions; (c) The SpliCons that produce the output patterns shown in (a) and (b); (d) The DOENet architecture. The first square shape represents intensity distributions input to the model. The following square shapes in the scheme are for extracted features through the DOENet. Each green square represents extracted features after a CNN layer with a ReLU activation function. Then, a max-pooling is performed, and size of extracted features is reduced to half of size of the input intensity distributions. Each blue square represents extracted features after a CNN layer with a ReLU activation function is performed. Then with a second max-pooling operation, size of extracted features is again reduced. Violet square indicates extracted features after a CNN layer with a ReLU activation function. The first red rectangle represents extracted features after flattening extracted features in violet square and utilizing a dense layer with a ReLU activation function. The next two red rectangles represent extracted features after a dense layer with a ReLU activation function. After a drop-out and a dense layer, features are presented with a black rectangle. Then reshaping and softmax activation operations are performed before correlating extracted features with SpliCons (see details of the DOENet architecture); (e) DOENet outputs: reconstructed SpliCons; (f) Optimized SpliCons with the hybrid algorithm; Calculated SOE distributions of the SpliCons shown in (f) for input wavelengths (g) at 1100 nm, (h) at 700 nm.}
  \label{dichromatic}
\end{figure*}

\section{Methods}

\subsection{Generation of data set}

We generate data sets for (i) concentrating dichromatic light source at two targets of the output plane with SpliCons, (ii) spectrally splitting the broadband light at two regions of the output plane with a broadband DOE, and (iii) spectrally splitting and spatially concentrating the broadband light at two target spots located at the output plane with a broadband SpliCon. For data set generation, we employed Fresnel-Kirchhoff diffraction integral to calculate an intensity distribution formed by a DOE/SpliCon at the input plane which is illuminated by a continuous plane wave. While optimizing a DOE/SpliCon we used the local search optimization algorithm. The decision-making logic of the local search optimization algorithm is the AND which improves fractional intensities at the targets at each wavelength/band simultaneously\cite{Yolalmaz2020}. Training data set generation takes 26 minutes per design wavelength on a PC, and that is a one-time effort.

Square-shaped DOEs/SpliCons are designed with a side length of \SI{200}{\micro\metre}. The DOEs/SpliCons consist of 1600 pixels with a \SI{5}{\micro\metre} pixel size. The DOEs/SpliCons have multilevel discrete pixels where thickness profile varies spatially between \SI{1}{\micro\metre} and \SI{8}{\micro\metre} with a \SI{1}{\micro\metre} step size. Design parameters of the DOEs/SpliCons are chosen based on what could practically be fabricated and be used in compact solar cell modules. The distance between a DOE/SpliCon and an output plane is \SI{350}{\micro\metre} to keep the diffraction pattern within the Fraunhofer field. For evaluating performance of a DOE/SpliCon, we use spectral optical efficiency (SOE) at a wavelength as a metric that defines percentage intensity at the wavelength on a target of the output plane.

We create our training data set while concentrating dichromatic light source at wavelengths of 700 nm and 1100 nm on two individual pixels at the output plane as seen in Fig. \ref{dichromatic}a-b. Given the fact that phase modulation of a longer wavelength requires a thicker medium for phase control, we chose to first control the longer wavelength limits of the two bands that are chosen between 400 nm - 700 nm and 701 nm - 1100 nm. After optimization of a SpliCon, the algorithm yields 12800 distinct SOE profiles (Fig. \ref{dichromatic}a-b) and SpliCons (Fig. \ref{dichromatic}c). Then, the same dichromatic light source is concentrated to 25 alternating target positions. The output plane with size of \SI{200}{\micro\metre} x \SI{200}{\micro\metre} is pixelated, and size of a pixel is \SI{20}{\micro\metre} x \SI{20}{\micro\metre} on the output plane.

Later we designed a broadband DOE to disperse the broadband light (400 nm - 1100 nm) into two bands that are the visible band (400 nm - 700 nm) and the short-IR band (701 nm - 1100 nm) on two regions at the output plane. The target region size is \SI{100}{\micro\metre} x \SI{200}{\micro\metre} for each band. Lastly, we optimized a broadband SpliCon to spectrally split the broadband light into the visible band and the short-IR band and simultaneously spatially concentrate on a target with a size of \SI{50}{\micro\metre} x \SI{100}{\micro\metre} for each band. The size of each pixel on the output plane is \SI{5}{\micro\metre} x \SI{5}{\micro\metre} for the broadband DOE/SpliCon. We use the bandwidth approach in our initial designs in order to minimize the computation time\cite{Yolalmaz2020}. The bandwidth approach reduces the number of design wavelengths from 701 with a 1 nm wavelength step to 54 with a 13 nm wavelength step using the DOE/SpliCon dimension and resolution that we employ here. During the optimization, we improve the mean SOE of each band simultaneously at each iteration, so we decrease degree of freedom in optimization for choosing proper DOE/SpliCon profile while keeping performance of the DOE/SpliCon same. Through the optimization of the broadband DOE/SpliCon, the algorithm yields 12800 distinct DOEs/SpliCons and corresponding SOE distributions as a data set for developing a deep learning model.

\subsection{Deep learning model}

Fresnel-Kirchhoff diffraction integral is an important tool to compute diffracted light pattern by a DOE/SpliCon. Tuning the topography of a DOE/SpliCon is a suitable problem that can be addressed using convolutional neural network (CNN) layers. Thus, in our deep learning model, we employed CNN layers to mimic relations between an intensity distribution of formed diffraction pattern and a corresponding DOE/SpliCon. The scheme of our deep learning model is presented in Fig. \ref{dichromatic}d that includes multiple CNN layers. The model has a sequence of 10 CNN layers each of which has 64 filters with a filter size 3 x 3, three max-pooling layers, and four fully connected layers. The CNN layers have the valid-padded convolutions. Between each CNN layer, ReLU activation function is accommodated to ignite nonlinearity which is also seen between a DOE/SpliCon and its intensity distribution. This activation function squashes the intensity distribution by performing a mathematical operation that gives input value if input value is greater than zero; otherwise, the ReLU gives zero. Then, down-sampling with max-pooling is performed to reduce the spatial size of the intensity distributions after 6th and 9th CNN layers. The number of parameters and computation load in the network is reduced after each max-pooling. Then, a flattening operation is performed to reshape extracted features. Later, three fully connected layers with 1600 units and three ReLU activation functions are used in our model which concatenate features from all previous layers. Then another fully connected layer with 12800 units is employed to make size of extracted features the same as size of the output data (the DOEs/SpliCons). The fully connected layers eliminate extracted information after each CNN layer to wash out by time due to an intensity distribution that passes through many layers in the deep learning model. The fully connected layers link to all extracted features, and they are prone to over-fitting. Here, we use a dropout with a factor of 0.2 to eliminate over-fitting during the training before the last fully connected layer. Lastly, we reshape size of extracted features into 40-by-40-by-8 to easily correlate them with the output data. We call our deep learning model diffractive optical element neural network (DOENet).

We normalized the intensity distributions obtained by the DOEs/SpliCons to arrange intensity values of the pixels between 0 and 1. The DOEs/SpliCons were converted to one-hot vectors with eight classes due to eight-level thickness variation of a DOE/SpliCon. We call Keras, Tensorflow, and Numpy libraries to access deep learning operations and mathematical tools. We used the ADAM optimizer to minimize categorical cross-entropy loss function over the training samples. Through searching weights of the model, we select a batch size of 100. Training of the model is completed in less than an hour using the Tensorflow library and an NVIDIA Quadro P5000 GPU. Once the training is completed, we test our model with validation data set which is 10\% of the input data set that is not part of the training set. The validation data set prevents over-fitting of the network model to the training set. Compared to the local search optimization algorithm design duration of a DOE/SpliCon for concentrating the dichromatic light via the DOENet decreases from 26 minutes to a few seconds after training the model. The decrease in design time is especially important for broadband light that typically takes 24 hours using the local search optimization algorithm. The DOENet speeds up the design of a broadband DOE/SpliCon thickness profile from 24 hours to a few seconds.

\subsection{The hybrid algorithm}
The flowchart of the hybrid algorithm is described for concentrating the dichromatic light source in Fig. \ref{dichromatic}. The hybrid algorithm operates as follows: first, a DOE/SpliCon is reconstructed for light intensity distributions at two design wavelengths by using the DOENet (Fig. \ref{dichromatic}e). Then, the reconstructed DOE/SpliCon is further redesigned with the local search optimization algorithm and Fresnel-Kirchhoff diffraction integral (Fig. \ref{dichromatic}f). During further optimization, the local search optimization algorithm increases SOEs at two design wavelengths simultaneously at each thickness change attempt. Further optimization of the DOE/SpliCon is carried out for only one iteration, and this iteration lasts 52 minutes. By using the further optimized DOE/SpliCon, we obtained the SOE distributions at two design wavelengths (Fig. \ref{dichromatic}g-h). The same flowchart is extended to design a broadband DOE/SpliCon with our hybrid algorithm.

\section{Results and discussion}

\subsection{Spectral Splitting and Concentration of Dichromatic Light Source}

\begin{figure}[!htb]
  \centering
  
    \includegraphics[width=85mm]{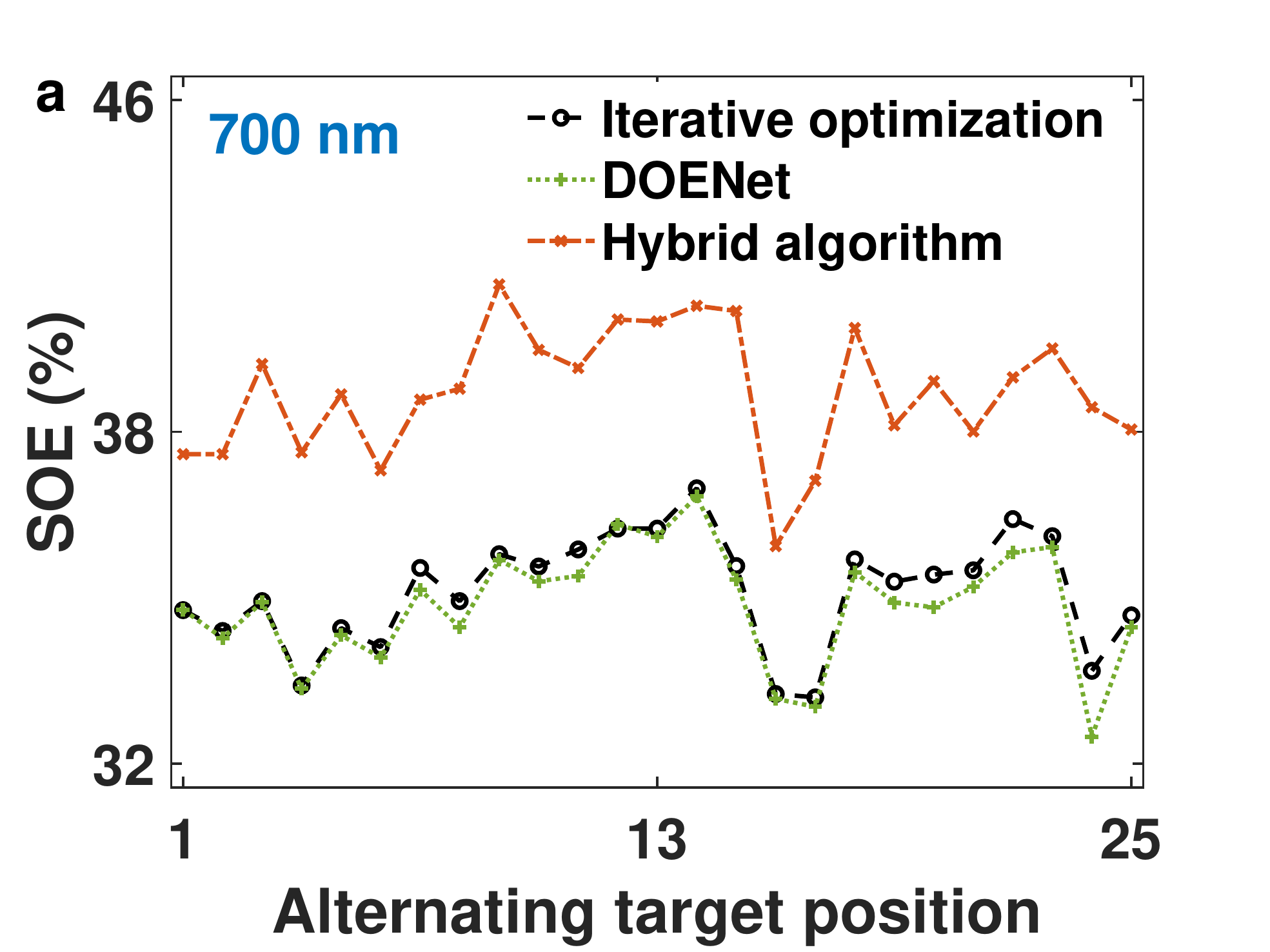}
    \includegraphics[width=85mm]{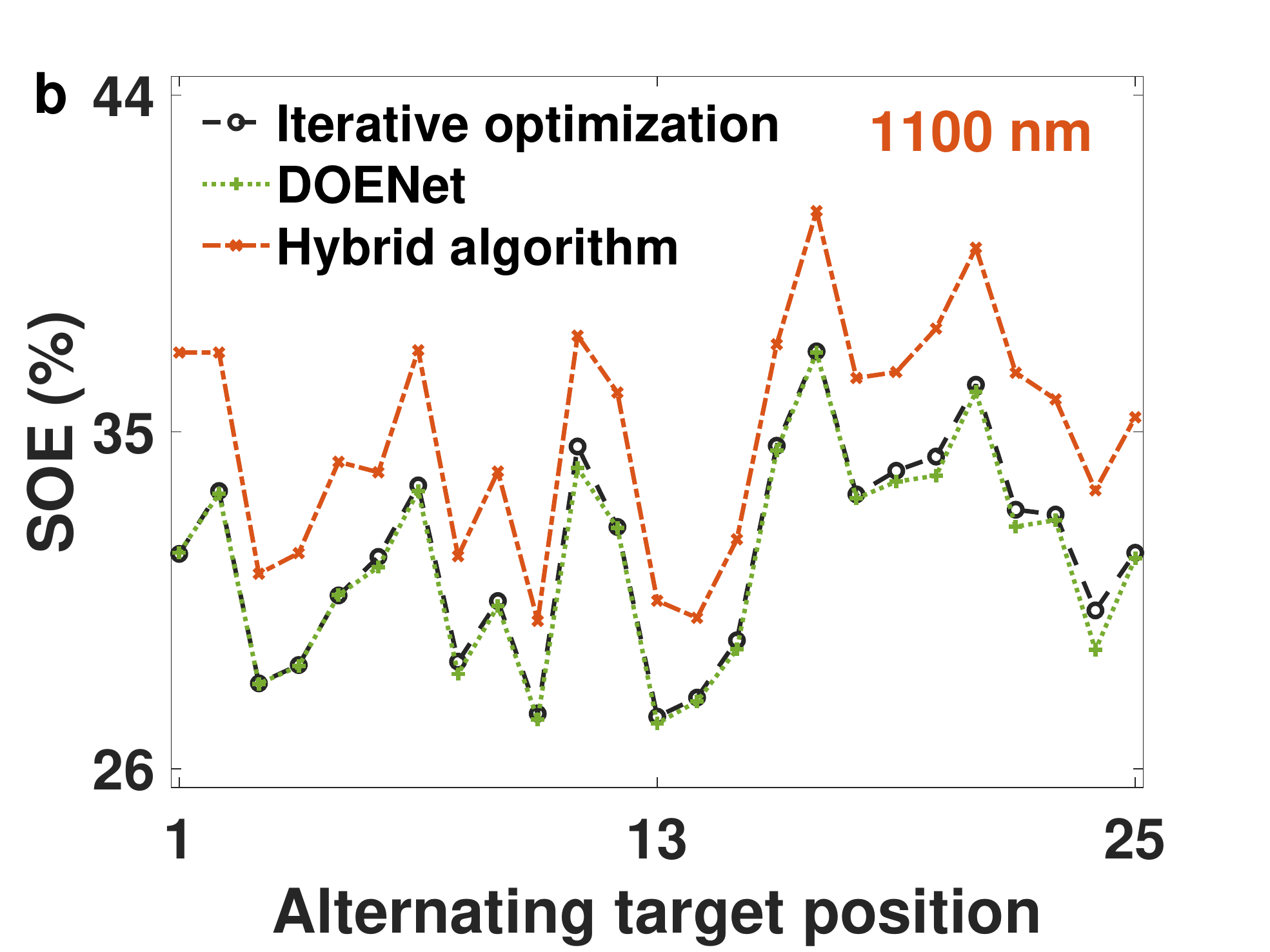}
    
  \caption{Concentrated light intensities in terms of SOE on alternating target positions computed for SpliCons which are obtained by the iterative algorithm, the DOENet, and the hybrid algorithm. (a) At a wavelength of 700 nm; (b) At a wavelength of 1100 nm. Lines are guides to the eye.}
  \label{dichromaticresult}
\end{figure}

The local search optimization algorithm arranges pixel thicknesses of the SpliCons, and at the end of optimization, the dichromatic light source is concentrated on two target pixels at the output plane. Later, the same light source is focused on alternating target positions with the same algorithm as seen in Fig. \ref{dichromatic}a-b. The SpliCon thickness profiles are presented in Fig. \ref{dichromatic}c. As a result of the local search optimization algorithm, we obtain a mean 35.6 $\pm$ 2.0\% SOE at 700 nm and a mean 31.8 $\pm$ 2.7\% SOE at 1100 nm on an individual pixel of the output plane (Fig. \ref{dichromaticresult}, Table I). We concentrate the dichromatic light source on a pixel with a 100 times smaller area than the total area of the output plane. As expected the highest enhancement we reach after optimization of a SpliCon, is 100. However, due to selected optimization parameters such as number of design wavelengths, the distance between the input and output planes, pixel size of the SpliCons, we obtained a mean 35.6 $\pm$ 2.0\ enhancement at 700 nm and a mean 31.8 $\pm$ 2.7\ enhancement at 1100 nm at the output plane.

\begin{table}[ht]
\caption{Average SOEs for the dichromatic light at 700 nm and 1100 nm which are calculated for the SpliCons obtained by the iterative optimization, the DOENet, and hybrid algorithm.}
\begin{center}
\begin{tabular}{ |p{2.6cm}|p{1.8cm}|p{1.8cm}|p{1.8cm}|  }

 \hline
 \textbf{Design}    & \textbf{SOE (\%)}   & \textbf{SOE (\%)} & \textbf{Design} \\
 \textbf{approach}    & \textbf{at 700 nm}   & \textbf{at 1100 nm} & \textbf{duration} \\
 \hline
 Iterative optimization  & 35.6 $\pm$ 2.0\%   &31.8 $\pm$ 2.7\% & 52 minutes\\
 \hline
 DOENet &  35.3 $\pm$ 1.7\%  & 31.6 $\pm$ 2.7\% & 2 seconds  \\
 \hline
 Hybrid algorithm &39.8 $\pm$ 1.4\% & 35.0 $\pm$ 3.0\% & 52 minutes\\
 
 \hline
\end{tabular}
\end{center}
\end{table}%

\begin{figure*}[!htb]
    \centering
    \includegraphics[width=180 mm]{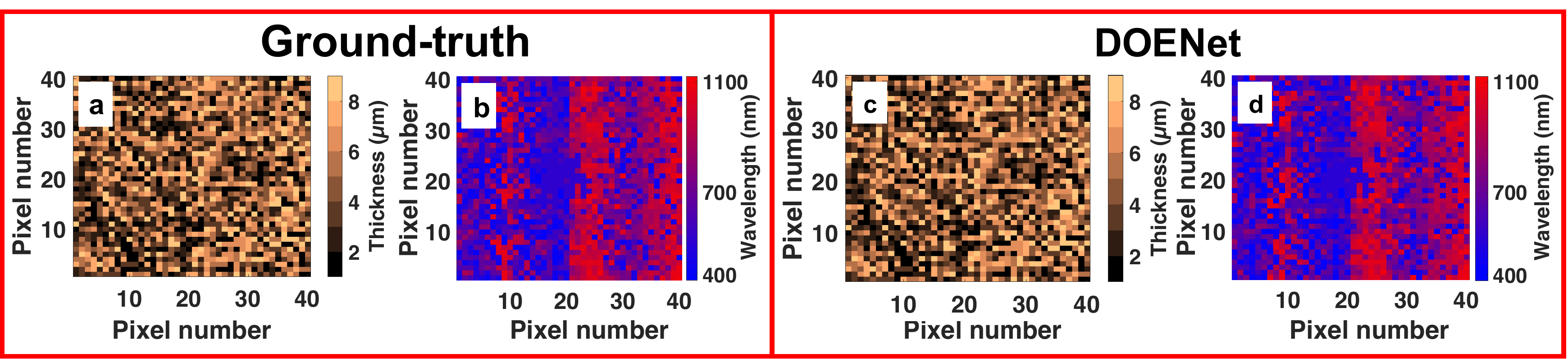}
    \caption{Spectral splitting of the broadband light into the visible and the short-IR bands. (a) Ground-truth thickness distribution of the DOE; (b) Wavelength distribution of the broadband light on the target; (c) Thickness distribution of the DOE obtained by the DOENet; (d) The wavelength distribution of the broadband light at the target achieved via DOEs that are designed with DOENet architecture.}
    \label{SpectralDOEWa}
\end{figure*}

After training the deep learning model DOENet (Fig. \ref{dichromatic}d) the model yields higher than 99\% accuracy within a training duration of 20 minutes for concentrating dichromatic light source at 700 nm and 1100 nm. Then, we reconstructed SpliCons using the model to concentrate dichromatic light source. The SpliCons obtained with the DOENet demonstrate similarities ranging between 99.9\% and 97.7\% with ground-truth SpliCons. Corresponding SOEs at 700 nm and 1100 nm computed with the reconstructed SpliCons for alternating target positions are presented in Fig. \ref{dichromaticresult}. As seen in this figure, the DOENet generates the SpliCons which show similar SOE results with the iterative optimization algorithm. We see an average 0.32\% error in SOE at 700 nm (Fig. \ref{dichromaticresult}a) and an average 0.21\% error in SOE value at 1100 nm (Fig. \ref{dichromaticresult}b). Moreover, with the DOENet we could design a SpliCon yielding a mean SOE of 35.3 $\pm$ 1.2 at 700 nm and a mean SOE of 31.6 $\pm$ 2.7 at 1100 nm for the individual targets (see Table I). These results highly overlap with mean SOEs received with the iterative optimization. Moreover, we could reconstruct a SpliCon by using the DOENet that concentrates light on 100 times smaller solar cell area, and we managed to confine light energy on the target area of \SI{20}{\micro\metre} x \SI{20}{\micro\metre} with at least 31.6 times enhancement compared to the case where no SpliCon is used.

Next, we re-optimized the reconstructed SpliCons with the local search optimization algorithm to boost performance of the SpliCons in terms of SOE. As seen in Fig. \ref{dichromaticresult} further optimization of the SpliCons results in higher SOEs for all alternating target positions. This optimization lasts 52 minutes on a PC and enhances concentrated light intensity up to 7.0\% SOE at 700 nm (Fig. \ref{dichromaticresult}a) and 5.4\% SOE at 1100 nm (Fig. \ref{dichromaticresult}b). We obtained an average 4.5 $\pm$ 0.9\% SOE increase at 700 nm (Fig. \ref{dichromaticresult}a) and an average 3.8 $\pm$ 0.7\% SOE increase at 1100 nm (see Fig. \ref{dichromaticresult}b and Table I). Compared to the case without a SpliCon, we reached up to an excess 41.1\% SOE. Considering the amount of energy received by the Sun, these excess increases that we achieve will lead to a significant rise in converted solar energy amount.

\begin{figure*}[!htb]
    \centering
    \includegraphics[width=180 mm]{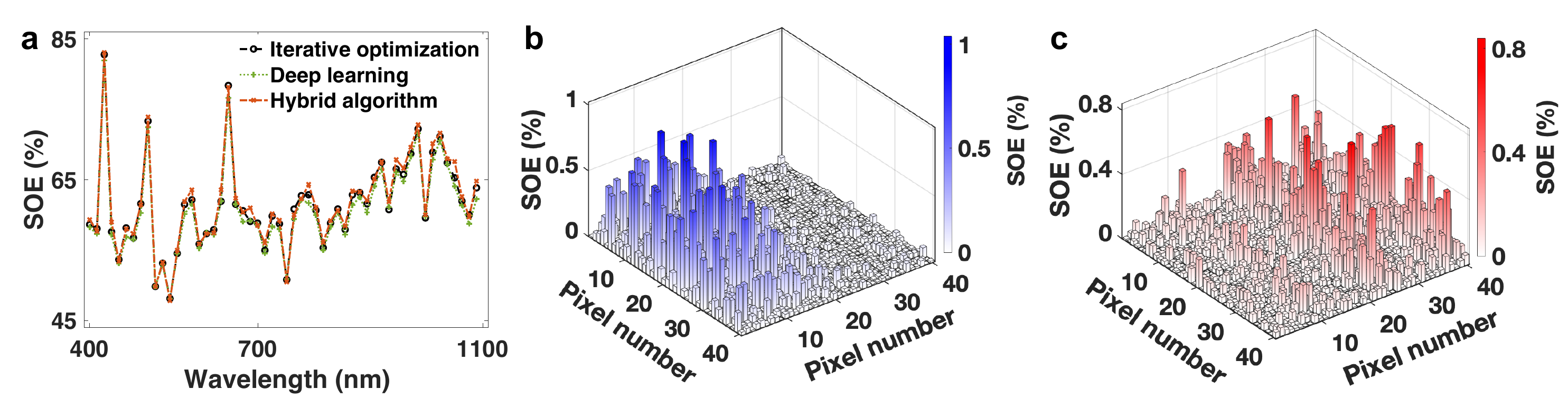}
    \caption{(a) The SOE spectrum for a DOE obtained with the iterative optimization, the DOENet, and the hybrid algorithm for spectral splitting the broadband light into the individual bands 400 nm - 700 nm (the visible band) and 701 nm - 1100 nm (the short-IR band). Lines are guides to the eye. The SOE distribution of the broadband light component calculated with the DOE obtained with the hybrid algorithm, (b) at 426 nm, (c) at 985 nm.}
    \label{spectral}
\end{figure*}

\subsection{Spectral Splitting of Broadband Light}

Later, we disperse the broadband light into two spectral regions: the visible band and the short-IR band. The thickness distribution of the DOE which shapes the incident light is shown in Fig. \ref{SpectralDOEWa}a. The dispersed visible band and short-IR band at two targets with sizes of \SI{100}{\micro\metre} x \SI{200}{\micro\metre} is shown in Fig. \ref{SpectralDOEWa}b. As observed here the visible band of the light source accumulates on the left half area of the target plane whereas the short IR band is concentrated on the right half of the target plane. The iterative optimization algorithm yields 60.0\% SOE in the visible band and 62.6\% SOE in the short-IR band for the targets (Fig. \ref{spectral}a, Table II).

\begin{table}[ht]
\caption{Mean SOEs for spectrally splitting the broadband light into the visible band and the short-IR band. The SOEs are calculated for the DOEs obtained by the iterative optimization, the DOENet, and the hybrid algorithm.}

\begin{center}
\begin{tabular}{ |p{2.6cm}|p{1.8cm}|p{2.0cm}| p{1.6cm}|  }

 \hline
 \textbf{Design}    & \textbf{SOE (\%)}   & \textbf{SOE (\%)} & \textbf{Design}\\
 \textbf{approach}    & \textbf{at visible}   & \textbf{at short-IR} & \textbf{duration}\\
 \hline
 Iterative optimization  & 60.0\%  & 62.6\% & 1 day \\
 \hline
 DOENet &  59.5\%  & 61.8\% & 2 seconds  \\
 \hline
 Hybrid algorithm & 60.4\% & 63.1\% & 1 day \\
 
 \hline
\end{tabular}
\end{center}
\end{table}%

The generated data during the iterative optimization is inserted into our deep learning model the DOENet as a training data set. This model gives 94.8\% accuracy with the training and validation data sets. Using the weights of the DOENet, we design a broadband DOE to spectrally split the broadband light which lasts 2 seconds (Fig. \ref{SpectralDOEWa}c). There is a 93.7\% correlation between the DOEs obtained by the iterative algorithm (ground-truth) and the DOENet. The wavelength distribution on the target with the DOE obtained by the DOENet is presented in Fig. \ref{SpectralDOEWa}d. As seen here, we reached a similar wavelength distribution at the target with that the iterative algorithm yields. By using the DOE designed by the DOENet, we presented SOE spectrum of the broadband DOE in Fig. \ref{spectral}a. Here, the DOENet results indicate good agreement with results attained by the iterative optimization. Quantitatively, we get a mean 59.5\% SOE at the visible band and a mean 61.8\% SOE at the short-IR band. Here, we see less than 1\% SOE difference in spectral splitting performance of the DOEs obtained via the iterative optimization algorithm and the DOENet. The results are summarized in Table II.

We further improve performance of the broadband DOE with the hybrid algorithm. At the end of the optimization with the hybrid algorithm, we reach a mean 60.4\% SOE at the visible band and a mean 63.1\% SOE at the short-IR band (Fig. \ref{spectral}a, Table II). Comparing the results of our hybrid algorithm with that of the DOENet we get 1.5\% and 2.1\% enhancements in performance of the DOE at the visible band and the short-IR band, respectively. In Fig. \ref{spectral}b-c we present distributions of SOE at wavelengths of 426 nm and 985 nm, which are obtained via the hybrid algorithm. As clearly visualized in Fig. \ref{spectral} the light source at the wavelength of 426 nm strongly generates signal on the left half of the target plane with 83.1\% cumulative SOE (Fig. \ref{spectral}b). In the meantime, the light at 985 nm strongly produces signal on the right half of the target plane with 72.8\% cumulative SOE (Fig. \ref{spectral}c).

\subsection{Spectral Splitting and Concentration of Broadband Light Source}

\begin{figure}[!htb]
  \centering
  
    \includegraphics[width=90mm]{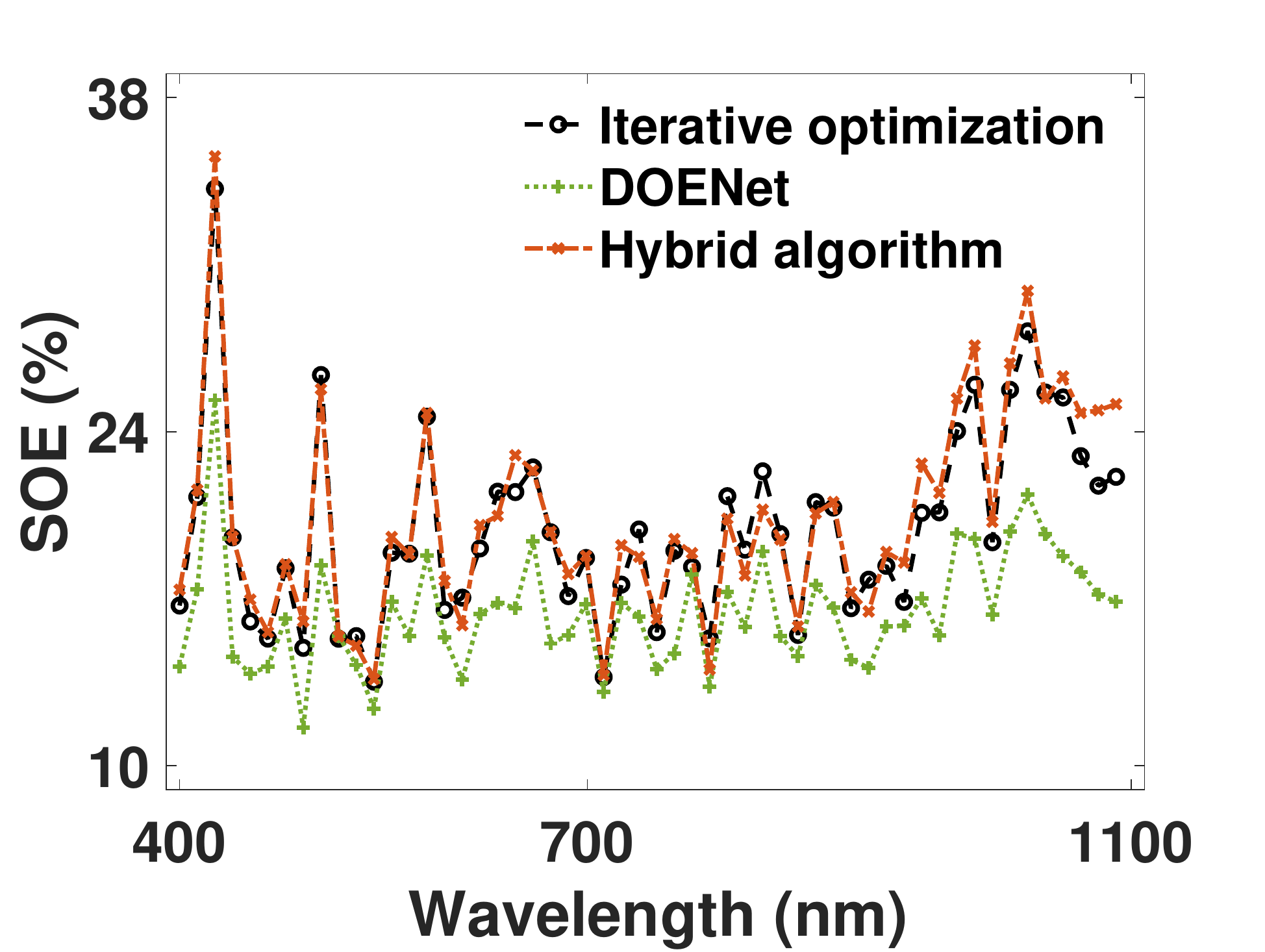} 
  \caption{The SOE spectra for spectrally splitting and spatially concentrating the broadband light 400 nm - 1100 nm into the visible band 400 nm - 700 nm and the short-IR band 701 nm - 1100 nm on a small area of the target with dimensions \SI{50}{\micro\metre} x \SI{100}{\micro\metre} for each band. The SOE spectra are computed for the SpliCons which are obtained with the iterative algorithm, the DOENet, and the hybrid algorithm. Lines are guides to the eye.}
  \label{broadsplicon}
\end{figure}

We first perform an iterative optimization to generate a broadband SpliCon using the local search algorithm to simultaneously spectrally split and spatially concentrate the broadband light on a small area at the target with dimensions \SI{50}{\micro\metre} x \SI{100}{\micro\metre} for each band. The SOE spectrum for the SpliCon is presented in Fig. \ref{broadsplicon}. The optimized SpliCon results in a mean 19.3\% SOE for the visible band and a mean 20.4\% SOE for the short-IR band (Table III). Compared to the lack of the SpliCon, we reach a 6.8\% excess SOE for the visible band and a 7.9\% excess SOE for the short-IR band with one-day SpliCon optimization. Using the data set acquired during the optimization of the broadband SpliCon we tune weights of the DOENet to speed up design process of a broadband SpliCon. This model gives 86.3\% accuracy with the training and validation data sets. Later, we reconstructed a broadband SpliCon using weights of the model to spectrally split and spatially concentrate the broadband light. There is a 70.7\% correlation in the SpliCons obtained by the iterative optimization and the DOENet. The SOE values at the design wavelengths of the SpliCon received via the DOENet are presented in Fig. \ref{broadsplicon}. As seen in this figure there is a high correlation between light intensity at the target obtained by the iterative optimization and the DOENet. Using the reconstructed SpliCon via the DOENet within a few seconds the SpliCon performs spectral splitting and spatially concentrating the broadband light with 16.0\% and 16.8\% mean SOEs for the visible band and the short-IR band, respectively (Table III).

\begin{table}[ht]
\caption{Mean SOEs for spectral splitting and spatially concentrating the broadband light into the visible band and the short-IR band. The SOEs are calculated for the SpliCons obtained by the iterative optimization, the DOENet, and the hybrid algorithm.}

\begin{center}
\begin{tabular}{ |p{2.6cm}|p{1.8cm}|p{2.0cm}| p{1.6cm}|  }

 \hline
 \textbf{Design}    & \textbf{SOE (\%)}   & \textbf{SOE (\%)} & \textbf{Design}\\
 \textbf{approach}    & \textbf{at visible}   & \textbf{at short-IR} & \textbf{duration}\\
 \hline
 Iterative optimization  & 19.3\%  & 20.4\% & 1 day \\
 \hline
 DOENet &  16.0\%  & 16.8\% & 2 seconds  \\
 \hline
 Hybrid algorithm & 19.6\% & 20.9\% & 1 day \\
 
 \hline
\end{tabular}
\end{center}
\end{table}%

\begin{figure}[!htb]
  \centering
  
    \includegraphics[width=80mm]{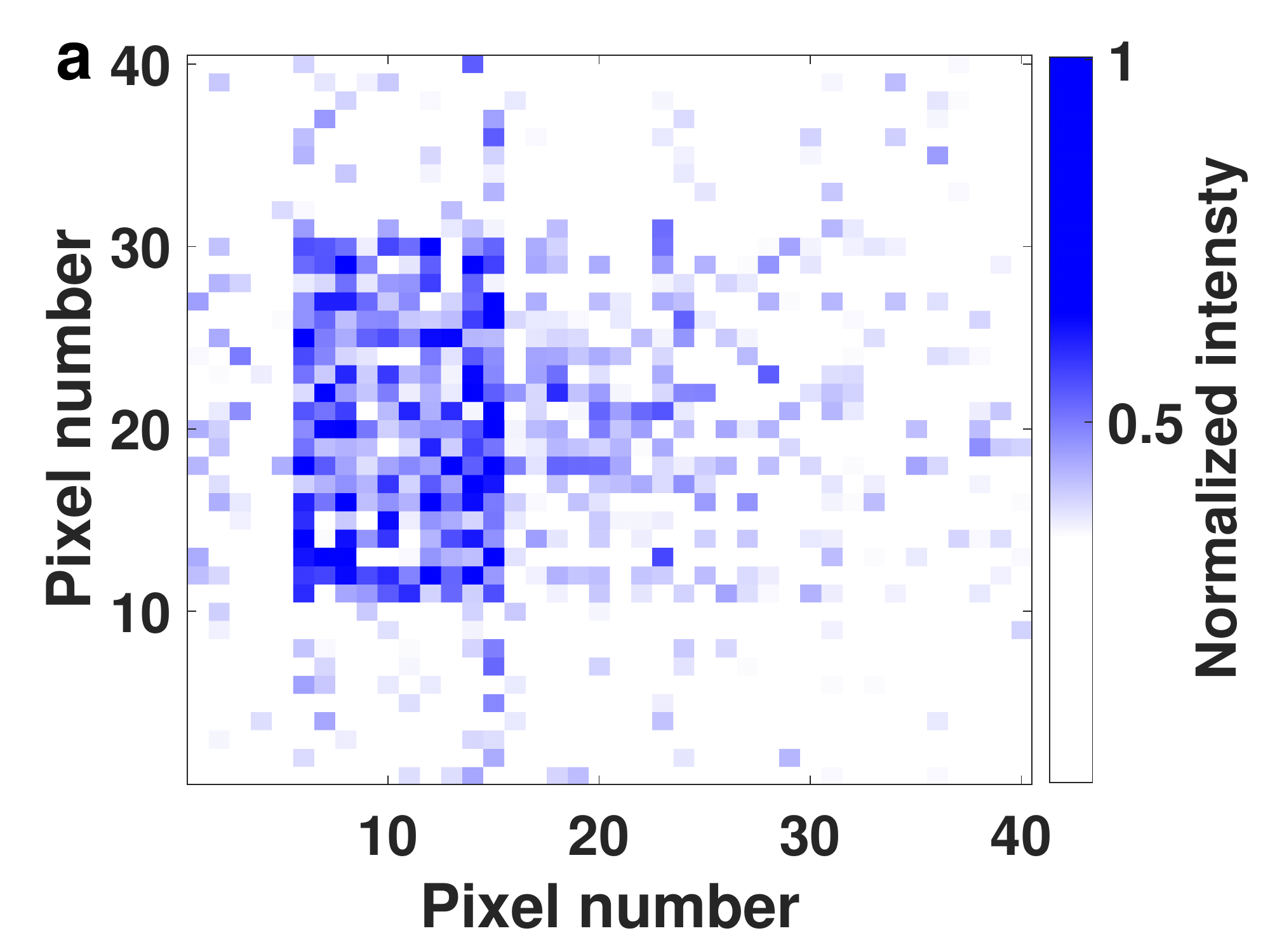}
    \includegraphics[width=80mm]{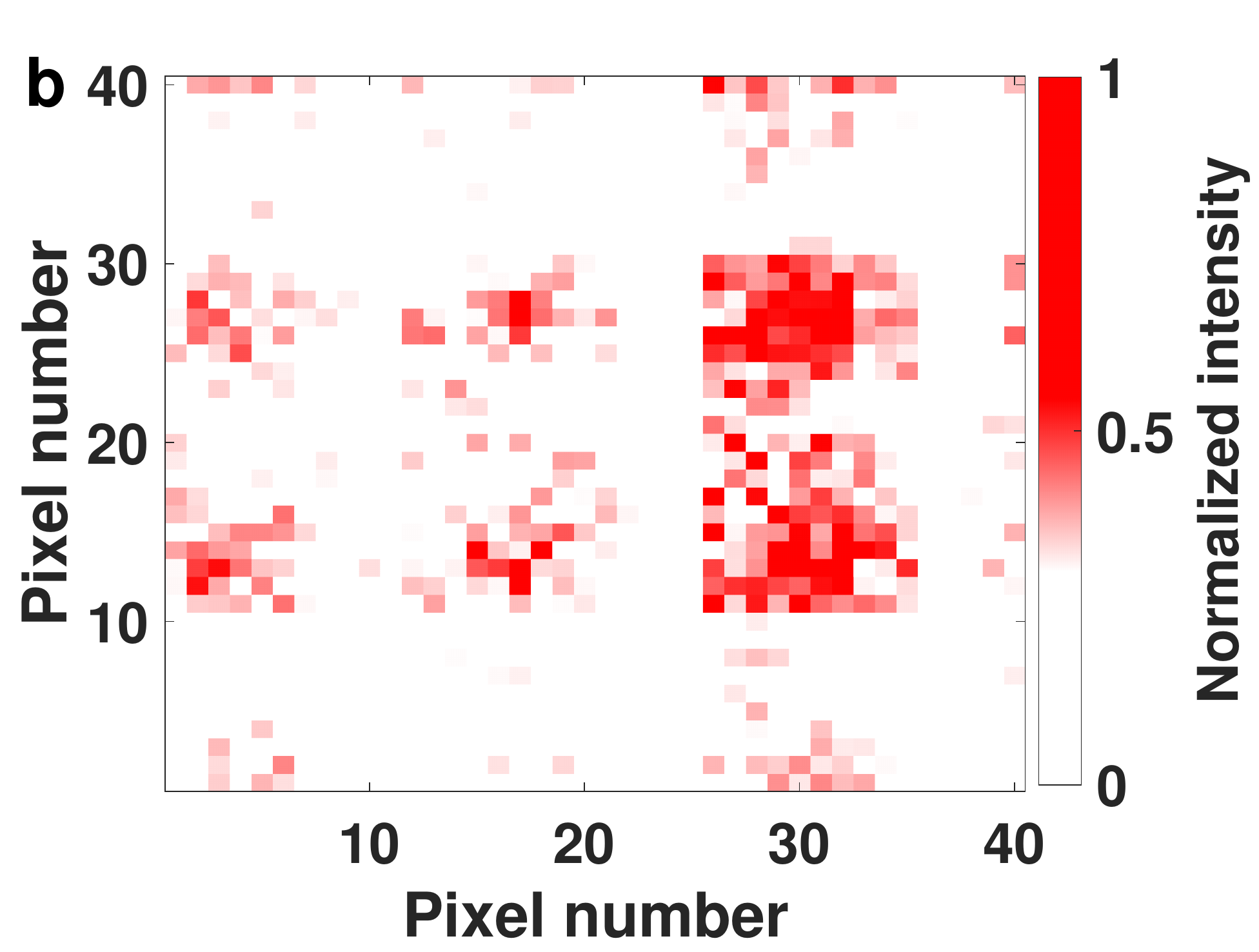}
    
  \caption{The normalized intensity distribution of the spectrally split and spatially concentrated broadband light obtained with the SpliCon using the hybrid algorithm. (a) The intensity distribution for (a) the visible band (400 nm - 700 nm) and (b) the short-IR band (701 nm - 1100 nm).}
  \label{splicondistribution}
\end{figure}

Next, we improve performance of the SpliCon with the hybrid algorithm. Starting with the SpliCon topography obtained via the DOENet we further optimize the SpliCon topography via the iterative search algorithm. Similar to earlier steps our hybrid algorithm again provides improved performance. Thereby, the robustness of our hybrid algorithm is proven for different cases that are addressed in this study. As shown in Fig. \ref{broadsplicon} the hybrid algorithm yields higher results in SOE spectrum while it significantly reduces the designing duration of the broadband SpliCon. The SpliCon designed via the hybrid algorithm shows a mean 19.6\% SOE at the visible band and a mean 20.9\% SOE at the short-IR band (Table III). In terms of enhancement, the SpliCon demonstrates 57\% and 67\% excess light concentrations at the visible band and the short-IR band at the targets, respectively. Our hybrid algorithm increases splitting and concentrating performance of the SpliCon by at least 22.4\% compared to the DOENet based SpliCon reconstruction. With the hybrid algorithm, we design a broadband SpliCon which yields a 7.1\% excess SOE at the visible band and an 8.4\% excess SOE at the short-IR band at a target area of \SI{50}{\micro\metre} x \SI{100}{\micro\metre} for each band.

The normalized intensity distribution at the output plane clearly indicates that the visible band of the broadband light is concentrated within the blue area in Fig. \ref{splicondistribution}a. Due to the fact that different light components of the broadband light show different phase shift through the same SpliCon, incoming light is directed outside the target area. With the hybrid algorithm, we limited to spread the intensity distribution outside the target. The blue area bears 19.6\% of the visible band intensity. The intensity distribution of the short-IR band is shown in Fig. \ref{splicondistribution}b. 20.9\% of the short-IR band is located within the red rectangular area. The diffraction nature of the light causes generation of diverse orders and leads to light spread outside of the target. In spite of that, we collect the short-IR portion of the broadband light with 20.9\% SOE on the area which is eight times smaller than the total area of the output plane.

\section{Conclusion}

Here, we present an exceptional design scheme of a DOE/SpliCon with our hybrid algorithm that not only improves performance of a DOE/SpliCon but also greatly reduces its optimization duration. Using the deep learning architecture the DOENet we achieve \textit{fast} and \textit{efficient} design of diffractive structures that perform concentration and spectral splitting of light. The DOENet we develop here yields up to 99\% accuracy in design of a DOE/SpliCon topography. With the presented hybrid approach, we spectrally split and spatially concentrate light using a SpliCon that provides up to a 41.1\% increase in SOE efficiency. Thank to the hybrid approach we reached an excess light intensity that helps in order to reduce our energy demand after harvesting much more solar energy with a DOE/SpliCon\cite{Mohammad2014, Mohammad2016, Kim2013}. Our hybrid design approach may enable one to improve spectral resolution of optical spectrometers including a DOE/SpliCon designed for high spectral splitting within a short design duration. The presented approach also is applicable to tune structural parameters of complex photonic devices operating across a wide parameter space which are functionalized for a variety of applications. We believe that using the DOENet, design of photonic structure may be up-scaled to a higher pixel number that is required to achieve improved control on broadband light.

\begin{acknowledgments}

This study is financially supported by The Scientific and Technological Research Council of Turkey (TUBITAK), grant no 118F075. Ph.D. study of Alim Yolalmaz is supported by TUBITAK, with a grant program of 2211-A. We thank Olca Orakçı for his help and suggestions about developing the DOENet.

\end{acknowledgments}

\bibliography{R-DL-DOE}

\end{document}